\documentclass[preprint]{ptephy_v1}%%%%%% to generate preprint number

\preprintnumber{KOBE-TH-22-01, KYUSHU-HET-233} %%% %%% Insert preprint number here

\usepackage{amsmath} %for dealing with mathematics,
\usepackage{amsthm} %for dealing with theorem environments,
\usepackage{url} %It provides better support for handling and breaking URLs.

\usepackage{stmaryrd}

\DeclareMathOperator{\tr}{tr}

\newcommand{\Slash}[1]{{\ooalign{\hfil/\hfil\crcr$#1$}}}
\numberwithin{equation}{section}
\allowdisplaybreaks

\begin{document}

\title{One-particle irreducible Wilson action in the gradient flow exact
renormalization group formalism}

%%%% To generate auto affiliation numbers please use \author{}\affil{} command

\author{Hidenori Sonoda}
\affil{Physics Department, Kobe University, Kobe 657-8501, Japan}

\author[2]{Hiroshi Suzuki}
\affil{Department of Physics, Kyushu University, 744 Motooka, Nishi-ku,
Fukuoka 819-0395, Japan}

%% \author{Insert second author name here}
%% \affil{Insert second author address here}

%% \author{Insert third author name here}
%% \author[3]{Insert fourth author name here} %%% Use optional bracket [3] to change the respective address
%% \affil{Insert third author address here}

%% \author{Insert last author name here\thanks{These authors contributed equally to this work}}
%% \affil{Insert last author address here}

%%% To include the collaborator name... Please use the command "\collaborator"
%%% For example: \collaborator{ATLAS Collaboration}

\begin{abstract}%
We define a one-particle irreducible (1PI) Wilson action in the gradient flow
exact renormalization group (GFERG) formalism as the Legendre transform of a
Wilson action. We consider quantum electrodynamics in particular, and show that
the GFERG flow equation preserves the invariance of the 1PI Wilson action
(excluding the gauge-fixing term) under the \emph{conventional\/} $U(1)$ gauge
transformation. This is in contrast to the invariance of the original Wilson
action under a modified $U(1)$ gauge transformation. The global chiral
transformation also takes the \emph{conventional\/} form for the 1PI Wilson
action. Despite the complexity of the GFERG flow equation, the conventional
form of the gauge and global chiral transformations may allow us to introduce a
non-perturbative Ansatz for gauge and chiral invariant 1PI Wilson actions.
\end{abstract}

\subjectindex{B05, B32}

\maketitle

\section{Introduction}
\label{sec:1}
The exact renormalization group (ERG)~\cite{Wilson:1973jj,Wegner:1972my,%
Wegner:1972ih,Morris:1993qb,Pawlowski:2005xe,Igarashi:2009tj,Rosten:2010vm,%
Dupuis:2020fhh} is a fundamental framework to investigate possible quantum
field theories even at a non-perturbative level. In a series of
papers~\cite{Sonoda:2020vut,Miyakawa:2021hcx,Miyakawa:2021wus}, we and a
collaborator have been developing a new formulation of ERG in gauge theory,
which we call the gradient flow exact renormalization group (GFERG). The basic
idea of GFERG comes from a formulation of ERG through the linear diffusion of
fields~\cite{Sonoda:2019ibh}. (See also~Refs.~\cite{Kagimura:2015via,%
Yamamura:2015kva,Aoki:2016ohw,Pawlowski:2017rhn,Makino:2018rys,Abe:2018zdc,%
Carosso:2018bmz,Carosso:2019qpb,Matsumoto:2020lha,Abe:2022smm} for related
investigations.) Using the non-linear diffusion equations introduced
in~Refs.~\cite{Narayanan:2006rf,Luscher:2009eq,Luscher:2010iy,Luscher:2011bx,%
Luscher:2013cpa} that preserve the gauge invariance, it might be possible to
generalize ERG to have manifest gauge invariance. This reasoning has led to
GFERG~\cite{Sonoda:2020vut,Miyakawa:2021hcx,Miyakawa:2021wus}.

Though it is possible to formulate gauge theory in the standard formulation of
ERG~\cite{Becchi:1996an, Ellwanger:1994iz,Bonini:1994dz,Bonini:1994kp,%
Reuter:1993kw,Reuter:1994sg}, the gauge transformation must be modified with
dependence on the Wilson action. Accordingly the gauge invariance is difficult
to implement as an Ansatz in any \emph{non-perturbative} but practical studies
of Wilson actions in gauge theory. (See
Refs.~\cite{Sonoda:2007dj,Igarashi:2021zml} for detailed studies of this issue
in perturbation theory.) Thus, if one aims at non-perturbative applications of
ERG in gauge theory, an ERG formulation with manifest gauge invariance is
highly desirable if not essential.\footnote{We refer the reader to
Refs.~\cite{Morris:1998kz,Morris:1999px,Morris:2000fs,Arnone:2005fb,%
Morris:2006in,Wetterich:2016ewc,Wetterich:2017aoy} for alternative manifestly
gauge invariant ERG formulations of gauge theory.}

Now, in non-perturbative applications of ERG, it has become conventional to
consider the one-particle irreducible (1PI) Wilson action~\cite{Nicoll:1977hi,%
Wetterich:1992yh,Wetterich:1993ne,Morris:1993qb,Bervillier:2008an} instead of
the original Wilson action~\cite{Polchinski:1983gv}. One reason for this is
that the 1PI Wilson action tends to have a simpler structure and is thus more
``economical''; see, for instance, Ref.~\cite{Ishikake:2005rk} for an
illustration on this point.

In this paper, with this simplicity in mind, we develop a formulation of the
1PI Wilson action as the Legendre transform of the Wilson action in GFERG. We
consider only the Abelian gauge theory, quantum electrodynamics (QED), with the
gauge fixing as developed in~Ref.~\cite{Miyakawa:2021wus}. It turns out that a
simple Legendre transformation well adopted in ERG works perfectly here. (Our
prescription corresponds to the choice $K_\Lambda(p)=e^{p^2/\Lambda^2}$
and~$k_\Lambda(p)=p^2/\Lambda^2$ in~Eq.~(23) of~Ref.~\cite{Igarashi:2016qdr}, up
to simple redefinitions of field variables.) We find it remarkable that the
GFERG equation preserves the invariance of the 1PI Wilson action (excluding the
gauge-fixing term) under the \emph{conventional\/} form of the $U(1)$ gauge
transformation. The global chiral transformation also takes the
\emph{conventional\/} form. In return, however, we must endure the complexity
of the GFERG equation satisfied by the 1PI Wilson action; up to the third-order
functional derivatives of the Legendre transformed variables with respect to
the original field variables are required. This is to be compared with the
first-order derivatives required in the standard ERG 1PI formalism.
Nevertheless, we believe that the simple form taken by the gauge
transformations of the 1PI Wilson action is a great advantage when we come to
introduce an Ansatz for the gauge invariant 1PI Wilson action in
non-perturbative studies of the renormalization group flows.

This paper is organized as follows: In~Sect.~\ref{sec:2}, we recapitulate the
main results of~Ref.~\cite{Miyakawa:2021wus} in terms of the original Wilson
action. We then generalize the condition of the chiral invariance to
accommodate the chiral anomaly and the associated topological effect.
In~Sect.~\ref{sec:3}, we introduce the Legendre transform that defines the 1PI
Wilson action. It is observed that the gauge and global chiral transformations
take the conventional form. The gauge invariance of the 1PI action, an a priori
known gauge-fixing term excluded, is preserved under our GFERG flow. The GFERG
equation for the 1PI action is derived in~Sect.~\ref{sec:4}. Its complexity is
the price to pay for the manifest gauge invariance. In~Sect.~5, we obtain the
1PI Wilson action explicitly up to the second order in the gauge coupling~$e$
by the Legendre transformation of the Wilson action, which has been calculated
in~Ref.~\cite{Miyakawa:2021wus}. We confirm that the 1PI Wilson action is
simpler than the original Wilson action, which contains products of 1PI parts
connected by short-distance propagators. We conclude the paper
in~Sect.~\ref{sec:5}.

Throughout the paper we work in the $D$-dimensional Euclidean space and
set~$D=4-\epsilon$. For the momentum integration, we use the abbreviation
\begin{equation}
   \int_p\equiv\int\frac{d^Dp}{(2\pi)^D}.
\label{eq:(1.1)}
\end{equation}

\section{Wilson action in GFERG}
\label{sec:2}
\subsection{GFERG equation}
\label{sec:2.1}
We recapitulate the essence of GFERG for QED formulated
in~Ref.~\cite{Miyakawa:2021wus}. We write all the expressions in dimensionless
variables, a convention suitable for the investigation of potential fixed
points.

The GFERG flow equation for the Wilson action~$S$ in
QED~\cite{Miyakawa:2021wus} can be expressed as\footnote{Here, we omit the
Faddeev--Popov ghost sector, because it decouples completely under the GFERG
flow~\cite{Miyakawa:2021wus}. In Appendix~\ref{sec:A}, we consider the 1PI
Wilson action containing the ghost sector, and show its BRST invariance. We
also omit the suffix~$\tau$ from~$S_\tau$ for simplicity.}
\begin{align}
   &\partial_\tau S
\notag\\
   &=-\int d^Dx\,
   e^{-S}\frac{\delta}{\delta A_\mu(x)}
   \Hat{s}
   \left(2\partial_{x'}^2+\frac{D-2}{2}+\gamma+x'\cdot\partial_{x'}
   \right)A_\mu(x')
   \Hat{s}^{-1}e^S
\notag\\
   &\qquad{}
   +\int d^Dx\,e^{-S}\tr
   \frac{\overrightarrow\delta}{\delta\Bar{\psi}(x)}
\notag\\
   &\qquad\qquad{}
   \times
   \Hat{s}
   \left[2\partial_{x'}^2+4ieA_\mu(x'')\partial_\mu^{x'}-2e^2A_\mu(x'')A_\mu(x''')
   +\frac{D-1}{2}+\gamma_F+x'\cdot\partial_{x'}
   \right]\Bar{\psi}(x')
\notag\\
   &\qquad\qquad\qquad{}
   \times\Hat{s}^{-1}e^S
\notag\\
   &\qquad{}
   +\int d^Dx\,
   \Hat{s}
   \tr
   \left[2\partial_{x'}^2-4ieA_\mu(x'')\partial_\mu^{x'}-2e^2A_\mu(x'')A_\mu(x''')
   +\frac{D-1}{2}+\gamma_F+x'\cdot\partial_{x'}
   \right]
\notag\\
   &\qquad\qquad\qquad\qquad{}
   \times\psi(x')\Hat{s}^{-1}e^S
   \frac{\overleftarrow\delta}{\delta\psi(x)}e^{-S},
\label{eq:(2.1)}
\end{align}
where $\tau$ parametrizes the logarithmic scale of the ERG transformation. The
differential operator~$\Hat{s}$, for which we coin the name ``scrambler'', is
defined by
\begin{align}
   \Hat{s}&\equiv
   \exp\left[\frac{1}{2}\int d^Dx\,
   \frac{\delta^2}{\delta A_\mu(x)\delta A_\mu(x)}\right]
   \exp\left[-i\int d^Dx\,
   \frac{\overrightarrow{\delta}}{\delta\psi(x)}
   \frac{\overrightarrow{\delta}}{\delta\Bar{\psi}(x)}\right].
\label{eq:(2.2)}
\end{align}
In Eq.~\eqref{eq:(2.1)} and expressions below, it is understood that primed
coordinates such as $x'$, $x''$, $x'''$, etc., are taken to~$x$ only after
functional differentiation. As elucidated in~Ref.~\cite{Miyakawa:2021wus}, this
``point-splitting prescription'' follows from a careful derivation of the GFERG
equation. In~Eq.~\eqref{eq:(2.1)}, $\gamma$ and~$\gamma_F$ are $\tau$-dependent
anomalous dimensions associated with the photon and electron fields,
respectively. These depend on the normalization condition adopted.

The GFERG equation~\eqref{eq:(2.1)} differs from the conventional ERG equation
for QED (see, for instance, Appendix~C of~Ref.~\cite{Miyakawa:2021wus}) simply
by the presence of terms containing the gauge coupling~$e$; it is those terms
that bring forth the fundamental property that the GFERG flow preserves
manifest gauge invariance. In~Ref.~\cite{Miyakawa:2021wus} we have introduced
\begin{equation}
   S_{\text{inv}}
   \equiv S
   +\frac{1}{2}\int d^Dx\, \partial_\mu A_\mu(x)
   \frac{1}
   {\xi E(-e^{-2\tau}\partial^2)e^{2\partial^2}-\partial^2}\partial_\nu A_\nu(x),
\label{eq:(2.3)}
\end{equation}
where $\xi$ is the gauge-fixing parameter, $E(x)$ is an arbitrary function
analytic at~$x=0$, and the gauge-fixing term is excluded from~$S$.\footnote{We
have generalized Eq.~(5.16) of~Ref.~\cite{Miyakawa:2021wus} by replacing $\xi$
by $\xi E(k^2e^{-2\tau})$ given in the momentum space. This is to be explained
below.} The GFERG equation~\eqref{eq:(2.1)} preserves the invariance
of~$S_{\text{inv}}$~\eqref{eq:(2.3)} under the modified gauge transformation
\begin{subequations}\label{eq:(2.4)}
\begin{align}
   \delta A_\mu(x)
   &=\frac{\xi E(-e^{-2\tau}\partial^2)e^{2\partial^2}-\partial^2}
   {\xi E(-e^{-2\tau}\partial^2)e^{2\partial^2}}
   \partial_\mu\chi(x),
\\
   \delta\psi(x)&=ie\chi(x)\psi(x),
\\
   \delta\Bar{\psi}(x)&=-ie\chi(x)\Bar{\psi}(x),
\end{align}
\end{subequations}
where $\chi(x)$ is an arbitrary infinitesimal function.

In~Ref.~\cite{Miyakawa:2021wus}, we have taken $E=1$ since we have restricted
our interest only to those $S$ parametrized by the gauge coupling~$e$, the
electron mass~$m$, and the gauge-fixing parameter~$\xi$. More generally, the
Faddeev--Popov ghost action must be given by
\begin{equation}
   S_{\text{ghost}}
   =\int d^Dx\,
   \Bar{c}(x)
   \frac{\partial^2}{E(-e^{2\tau}\partial^2)e^{2\partial^2}-\partial^2}c(x)
   =\int_k\,\Bar{c}(-k)\frac{-k^2}{E(k^2e^{-2\tau})e^{-2k^2}+k^2}c(k).
\label{eq:(2.5)}
\end{equation}
Then, repeating the argument in~Ref.~\cite{Miyakawa:2021wus} we obtain
Eqs.~\eqref{eq:(2.3)} and~\eqref{eq:(2.4)}.

We note in passing that the invariance of~$S_{\text{inv}}$~\eqref{eq:(2.3)}
under~Eq.~\eqref{eq:(2.4)} is equivalent to
\begin{equation}
   \delta S=\int d^Dx\,
   A_\mu(x)\frac{\partial^2}{\xi E(-e^{-2\tau}\partial^2)e^{2\partial^2}}
   \partial_\mu\chi(x).
\label{eq:(2.6)}
\end{equation}
This is to be used later.

Finally, the $\tau$-dependence of the gauge parameter in the above is given by
the anomalous dimension~$\gamma$ as~\cite{Miyakawa:2021wus}
\begin{equation}
   \partial_\tau e=\left(\frac{\epsilon}{2}-\gamma\right)e,
\label{eq:(2.7)}
\end{equation}
and that of the gauge-fixing parameter by
\begin{equation}
   \partial_\tau\xi=2\gamma\xi.
\label{eq:(2.8)}
\end{equation}
The relation~\eqref{eq:(2.7)} that corresponds to the Ward identity $Z_1=Z_3$
in the conventional formulation follows naturally from the definition of the
gauge coupling~$e$ in this formulation. Equation~\eqref{eq:(2.8)} follows from
the underlying BRST symmetry preserved by the GFERG equation.

\subsection{Chiral symmetry}
\label{sec:2.2}
In~Ref.~\cite{Miyakawa:2021hcx}, it was shown that the GFERG
equation~\eqref{eq:(2.1)} is consistent with a modified form of the chiral
symmetry \`a la Ginsparg--Wilson~\cite{Ginsparg:1981bj}. Writing the
differential operator that generates the conventional global chiral
transformation as
\begin{equation}
   \Hat{\gamma}_5
   \equiv-\int d^Dx\,
   \left[
   \frac{\overrightarrow{\delta}}{\delta\psi(x)}\gamma_5\psi(x')
   +\frac{\overrightarrow{\delta}}{\delta\Bar{\psi}(x)}\Bar{\psi}(x')\gamma_5
   \right],
\label{eq:(2.9)}
\end{equation}
we define the modified global chiral transformation by
\begin{equation}
   \Hat{{\mit\Gamma}}_5\equiv\Hat{s}\Hat{\gamma}_5\Hat{s}^{-1},
\label{eq:(2.10)}
\end{equation}
where $\Hat{s}$ is the scrambler defined by~Eq.~\eqref{eq:(2.2)}. Then the
modified chiral invariance of the Wilson action~$S$ is given as the vanishing
of $\Hat{\mit\Gamma}_5$ acting on $S$:
\begin{align}
   e^{-S}\Hat{{\mit\Gamma}}_5e^S
   &=\int d^Dx\,
   \left[
   S\frac{\overleftarrow{\delta}}{\delta\psi(x)}\gamma_5\psi(x)
   +\Bar{\psi}(x)\gamma_5\frac{\overrightarrow{\delta}}{\delta\Bar{\psi}(x)}S
   +2iS\frac{\overleftarrow{\delta}}{\delta\psi(x)}\gamma_5
   \frac{\overrightarrow{\delta}}{\delta\Bar{\psi}(x)}S
   \right]
\notag\\
   &\qquad{}
   +\int d^Dx\,
   \tr\left[
   -2i\gamma_5\frac{\overrightarrow{\delta}}{\delta\Bar{\psi}(x')}S
   \frac{\overleftarrow{\delta}}{\delta\psi(x)}
   \right]
\notag\\
   &=0.
\label{eq:(2.11)}
\end{align}
This constrains the Wilson action. If we assume that $S$ is bilinear in fermion
fields, Eq.~\eqref{eq:(2.11)} is nothing but the Ginsparg--Wilson
relation~\cite{Ginsparg:1981bj}.

Equation~\eqref{eq:(2.11)} is an ``operator equation'' preserved under the
GFERG equation~\eqref{eq:(2.1)}. To explain what we mean, we express the global
chiral transformation in terms of the so-called composite
operators~\cite{Becchi:1996an,Igarashi:2009tj}. We first introduce composite
operators corresponding to the elementary fields $A_\mu(x)$, $\psi(x)$,
and~$\Bar{\psi}(x)$ by
\begin{subequations}\label{eq:(2.12)}
\begin{align}
   \mathcal{A}_\mu(x)
   &\equiv A_\mu(x)+\frac{\delta S}{\delta A_\mu(x)}
   =e^{-S}\Hat{s}A_\mu(x)\Hat{s}^{-1}e^S,
\label{eq:(2.12a)}\\
   \Psi(x)
   &\equiv\psi(x)+i\frac{\overrightarrow{\delta}}{\delta\Bar{\psi}(x)}S
   =e^{-S}\Hat{s}\psi(x)\Hat{s}^{-1}e^S,
\label{eq:(2.12b)}\\
   \Bar{\Psi}(x)
   &\equiv\Bar{\psi}(x)+iS\frac{\overleftarrow{\delta}}{\delta\psi(x)}
   =e^{-S}\Hat{s}\Bar{\psi}(x)\Hat{s}^{-1}e^S,
\label{eq:(2.12c)}
\end{align}
\end{subequations}
where $\Hat{s}$ is the scrambler~\eqref{eq:(2.2)}. When inserted into the
functional integral, each of the above plays the role of an elementary field in
the \emph{modified\/} correlation function~\cite{Sonoda:2015bla}.\footnote{The
modified correlation function is defined by
\begin{equation}
   \left\langle\!\left\langle
   \varphi(x_1)\dotsb\varphi(x_n)\right\rangle\!\right\rangle
   \equiv
   \int d\mu\,e^S\Hat{s}^{-1}\varphi(x_1)\dotsb\varphi(x_n),
\label{eq:(2.13)}
\end{equation}
where $d\mu$ is the measure of functional integration over the elementary
field~$\varphi$, and $\Hat{s}^{-1}$ is the inverse of the
scrambler~\eqref{eq:(2.2)}. The modified correlation function first introduced
in~Ref.~\cite{Sonoda:2015bla} differs from~Eq.~\eqref{eq:(2.13)} by the
application of inverse diffusion, and exhibits a simple scaling behavior absent
in the ordinary correlation function.} For instance,
for~Eq.~\eqref{eq:(2.12a)}, we obtain
\begin{align}
   &\int d\mu\,
   e^S\mathcal{A}_\mu(x)\Hat{s}^{-1}
   \left[\psi(x_1)\dotsb\psi(x_n)\Bar{\psi}(y_1)\dotsb\Bar{\psi}(y_n)\right]
\notag\\
   &=\int d\mu\,\hat{s}A_\mu(x)\hat{s}^{-1}e^S
   \cdot\hat{s}^{-1}
   \left[\psi(x_1)\dotsb\psi(x_n)\bar{\psi}(y_1)\dotsb\bar{\psi}(y_n)\right]
\notag\\
   &=\int d\mu\,e^S\cdot\hat{s}^{-1}
   \left\{A_\mu(x)\hat{s}\hat{s}^{-1}
   \left[\psi(x_1)\dotsb\psi(x_n)\bar{\psi}(y_1)\dotsb\bar{\psi}(y_n)
   \right]\right\}
\notag\\
   &=\int d\mu\,e^S\cdot\hat{s}^{-1}
   \left[A_\mu(x)\psi(x_1)\dotsb\psi(x_n)\bar{\psi}(y_1)\dotsb\bar{\psi}(y_n)
   \right]
\notag\\
   &=\left\langle\!\left\langle
   A_\mu(x)\psi(x_1)\dotsb\psi(x_n)\Bar{\psi}(y_1)\dotsb\Bar{\psi}(y_n)
   \right\rangle\!\right\rangle.
\label{eq:(2.14)}
\end{align}

Using the composite operators in~Eq.~\eqref{eq:(2.12)}, we can introduce a
composite operator
\begin{equation}
   Q_5\equiv-e^{-S}\int d^Dx\,
   \tr\left\{
   \left[e^S\gamma_5\Psi(x')\right]
   \frac{\overleftarrow{\delta}}{\delta\psi(x)}
   +\frac{\overrightarrow{\delta}}{\delta\Bar{\psi}(x)}
   \left[\Bar{\Psi}(x')\gamma_5e^S\right]
   \right\}.
\label{eq:(2.15)}
\end{equation}
Using Eqs.~\eqref{eq:(2.12b)} and~\eqref{eq:(2.12c)}, we find
\begin{equation}
   Q_5=e^{-S}\Hat{\mit\Gamma}_5e^S.
\label{eq:(2.16)}
\end{equation}
$Q_5$ is a particular example of the equation-of-motion composite
operator~\cite{Becchi:1996an,Igarashi:2009tj}. The equation-of-motion composite
operator, when inserted into the functional integral, induces a transformation
(such as the field shift) one by one on each field in the modified correlation
function. The composite operator~$Q_5$ defined by~Eq.~\eqref{eq:(2.15)}
generates the global chiral transformation:
\begin{align}
   &\int d\mu\,e^SQ_5\Hat{s}^{-1}
   \left[\psi(x_1)\dotsb\psi(x_n)\Bar{\psi}(y_1)\dotsb\Bar{\psi}(y_n)\right]
\notag\\
   &=-\int d\mu\,e^S\Hat{s}^{-1}
   \int d^Dx\,
   \left[
   \gamma_5\psi(x')\frac{\overrightarrow{\delta}}{\delta\psi(x)}
   +\Bar{\psi}(x')\gamma_5\frac{\overrightarrow{\delta}}{\delta\Bar{\psi}(x)}
   \right]
   \left[
   \psi(x_1)\dotsb\psi(x_n)\Bar{\psi}(y_1)\dotsb\Bar{\psi}(y_n)\right]
\notag\\
   &=-\sum_{i=1}^n
   \left\langle\!\left\langle
   \psi(x_1)\dotsb\gamma_5\psi(x_i)\dotsb\psi(x_n)
   \Bar{\psi}(y_1)\dotsb\Bar{\psi}(y_n)
   \right\rangle\!\right\rangle
\notag\\
   &\qquad\qquad{}-\sum_{i=1}^n
   \left\langle\!\left\langle
   \psi(x_1)\dotsb\psi(x_n)
   \Bar{\psi}(y_1)\dotsb\Bar{\psi}(y_i)\gamma_5\dotsb\Bar{\psi}(y_n)
   \right\rangle\!\right\rangle.
\label{eq:(2.17)}
\end{align}

Now, $Q_5$ being a composite operator of scale dimension zero, $S+\eta Q_5$
satisfies the same GFERG equation~\eqref{eq:(2.1)} as~$S$ up to the first order
in the constant~$\eta$. In other words the $\tau$-dependence of $Q_5$ is linear
in~$Q_5$:
\begin{align}
   &\partial_\tau Q_5
\notag\\
   &=\left[
   -\int d^Dx\,
   \frac{\delta}{\delta A_\mu(x)}
   \Hat{s}
   \left(2\partial_{x'}^2+\frac{D-2}{2}+\gamma+x'\cdot\partial_{x'}
   \right)A_\mu(x')
   \Hat{s}^{-1},Q_5\right]
\notag\\
   &\qquad{}
   +\Biggl[
   \int d^Dx\,e^{-S}\tr
   \frac{\overrightarrow\delta}{\delta\Bar{\psi}(x)}
\notag\\
   &\qquad\qquad{}
   \times
   \Hat{s}
   \left[2\partial_{x'}^2+4ieA_\mu(x'')\partial_\mu^{x'}-2e^2A_\mu(x'')A_\mu(x''')
   +\frac{D-1}{2}+\gamma_F+x'\cdot\partial_{x'}
   \right]\Bar{\psi}(x')\Hat{s}^{-1},Q_5\Biggr]
\notag\\
   &\qquad{}
   +\Biggl[Q_5,
   \int d^Dx\,
   \Hat{s}
   \tr
   \left[2\partial_{x'}^2-4ieA_\mu(x'')\partial_\mu^{x'}-2e^2A_\mu(x'')A_\mu(x''')
   +\frac{D-1}{2}+\gamma_F+x'\cdot\partial_{x'}
   \right]
\notag\\
   &\qquad\qquad\qquad\qquad{}
   \times\psi(x')\Hat{s}^{-1}
   \frac{\overleftarrow\delta}{\delta\psi(x)}\Biggr],
\label{eq:(2.18)}
\end{align}
where the square brackets denote commutators. If
\begin{equation}
   Q_5=0
\label{eq:(2.19)}
\end{equation}
at some~$\tau$, this holds for any~$\tau$. This is what we meant by the
``operator equation'' at the beginning of the paragraph
below~Eq.~\eqref{eq:(2.11)}.

\section{1PI action}
\label{sec:3}
\subsection{Legendre transformation}
\label{sec:3.1}
We now define a 1PI Wilson action~$\mit\Gamma$ from the original Wilson
action~$S$ by the following Legendre transformation:
\begin{align}
   &{\mit\Gamma}[\mathcal{A}_\mu,\Psi,\Bar{\Psi}]
   -\frac{1}{2}\int d^Dx\,\mathcal{A}_\mu(x)\mathcal{A}_\mu(x)
   +i\int d^Dx\,\Bar{\Psi}(x)\Psi(x)
\notag\\
   &\equiv
   S[A_\mu,\psi,\Bar{\psi}]
   +\frac{1}{2}\int d^Dx\,A_\mu(x)A_\mu(x)
   -i\int d^Dx\,\Bar{\psi}(x)\psi(x)
\notag\\
   &\qquad{}
   -\int d^Dx\,\mathcal{A}_\mu(x)A_\mu(x)
   +i\int d^Dx\,\left[
   \Bar{\Psi}(x)\psi(x)+\Bar{\psi}(x)\Psi(x)
   \right],
\label{eq:(3.1)}
\end{align}
where the field variables $\mathcal{A}_\mu$, $\Psi$, and~$\Bar{\Psi}$ are
defined by~Eq.~\eqref{eq:(2.12)}. We can regard $\mit\Gamma$ as a functional of
these new variables. As is usual with the Legendre transformation, we have
relations ``dual'' to~Eq.~\eqref{eq:(2.12)}:
\begin{subequations}\label{eq:(3.2)}
\begin{align}
   \frac{\delta{\mit\Gamma}}{\delta\mathcal{A}_\mu(x)}-\mathcal{A}_\mu(x)
   &=-A_\mu(x),
\\
   i\frac{\overrightarrow{\delta}}{\delta\Bar{\Psi}(x)}{\mit\Gamma}-\Psi(x)
   &=-\psi(x),
\\
   i{\mit\Gamma}\frac{\overleftarrow{\delta}}{\delta\Psi(x)}-\Bar{\Psi}(x)
   &=-\Bar{\psi}(x).
\end{align}
\end{subequations}
These dual relations can also be summarized as
\begin{subequations}\label{eq:(3.3)}
\begin{align}
   \frac{\delta S}{\delta A_\mu(x)}
   &=\mathcal{A}_\mu(x)-A_\mu(x)
   =\frac{\delta{\mit\Gamma}}{\delta\mathcal{A}_\mu(x)},
\\
   i\frac{\overrightarrow{\delta}}{\delta\Bar{\psi}(x)}S
   &=\Psi(x)-\psi(x)
   =i\frac{\overrightarrow{\delta}}{\delta\Bar{\Psi}(x)}{\mit\Gamma},
\\
   iS\frac{\overleftarrow{\delta}}{\delta\psi(x)}
   &=\Bar{\Psi}(x)-\Bar{\psi}(x)
   =i{\mit\Gamma}\frac{\overleftarrow{\delta}}{\delta\Psi(x)}.
\end{align}
\end{subequations}

\subsection{Gauge transformation and invariance}
\label{sec:3.2}
We wish to find how $\mit\Gamma$ transforms under the gauge
transformation~\eqref{eq:(2.4)}. Using Eq.~\eqref{eq:(2.6)}, we obtain
\begin{subequations}\label{eq:(3.4)}
\begin{align}
   \delta\mathcal{A}_\mu(x)
   &=\delta A_\mu(x)+\frac{\delta}{\delta A_\mu(x)}\delta S
\notag\\
   &=\frac{\xi E(-e^{-2\tau}\partial^2)e^{2\partial^2}-\partial^2}
   {\xi E(-e^{-2\tau}\partial^2)e^{2\partial^2}}\partial_\mu\chi(x)
   +\frac{\partial^2}{\xi E(-e^{-2\tau}\partial^2)e^{2\partial^2}}
   \partial_\mu\chi(x)
\notag\\
   &=\partial_\mu\chi(x).
\end{align}
The transformation of $\Psi$ and~$\bar{\Psi}$ remains the same as $\psi$
and~$\bar{\psi}$:
\begin{align}
   \delta\Psi(x)&=ie\chi(x)\Psi(x),
\\
   \delta\Bar{\Psi}(x)&=-ie\chi(x)\Bar{\Psi}(x).
\end{align}
\end{subequations}
Thus, the U(1) gauge transformation of $\mathcal{A}_\mu(x)$, $\Psi(x)$,
and~$\Bar{\Psi}(x)$ takes the \emph{conventional\/} form.

Moreover, from~Eq.~\eqref{eq:(3.1)} we obtain
\begin{align}
   \delta{\mit\Gamma}
   &=\int d^Dx\,
   \mathcal{A}_\mu(x)\delta\mathcal{A}_\mu(x)
   +\delta S
   +\int d^Dx\,A_\mu(x)\delta A_\mu(x)
   -\int d^Dx\,\delta\left[\mathcal{A}_\mu(x)A_\mu(x)\right]
\notag\\
   &=-\int d^Dx\,
   \partial^2\chi(x)
   \frac{1}{\xi E(-e^{-2\tau}\partial^2)e^{2\partial^2}}
   \partial_\mu\mathcal{A}_\mu(x),
\label{eq:(3.5)}
\end{align}
where we have used Eqs.~\eqref{eq:(2.4)}, \eqref{eq:(3.4)},
and~\eqref{eq:(2.6)}. This shows that the combination
\begin{equation}
   {\mit\Gamma}_{\text{inv}}
   \equiv{\mit\Gamma}
   +\frac{1}{2}\int d^Dx\,
   e^{-\partial^2}\partial_\mu\mathcal{A}_\mu(x)
   \cdot 
   \frac{1}{\xi E(-e^{-2\tau}\partial^2)}
   \cdot
   e^{-\partial^2}\partial_\nu\mathcal{A}_\nu(x)
\label{eq:(3.6)}
\end{equation}
is invariant under~Eq.~\eqref{eq:(3.4)}, and it is preserved by the GFERG
equation. We believe that the conventional gauge invariance
of~${\mit\Gamma}_{\text{inv}}$ is a remarkable property of our 1PI formulation.
We expect this to facilitate the search for non-trivial fixed-point Wilson
actions, where some kind of truncation is inevitable in practice. It is
essential that the truncation keeps a gauge invariance that is under our
control.

\subsection{Global chiral transformation and symmetry}
\label{sec:3.3}
Using Eq.~\eqref{eq:(3.3)}, we can rewrite~$Q_5$~\eqref{eq:(2.15)} in terms of
the 1PI action~$\mit\Gamma$:
\begin{align}
   Q_5&=\int d^Dx\,
   \left[
   {\mit\Gamma}\frac{\overleftarrow{\delta}}{\delta\Psi(x)}
   \gamma_5\Psi(x')
   +\Bar{\Psi}(x')\gamma_5
   \frac{\overrightarrow{\delta}}{\delta\Bar{\Psi}(x)}{\mit\Gamma}
   \right]
\notag\\
   &\qquad{}
   -\int d^Dx\,
   \tr\left[
   \gamma_5\Psi(x')\frac{\overleftarrow{\delta}}{\delta\psi(x)}
   +\frac{\overrightarrow{\delta}}{\delta\Bar{\psi}(x)}\Bar{\Psi}(x')\gamma_5
   \right].
\label{eq:(3.7)}
\end{align}
Note that the first integral on the right-hand side is the variation of the 1PI
action under the \emph{conventional\/} global chiral transformation
\begin{equation}
   \delta\Psi(x)=\alpha\gamma_5\Psi(x),\qquad
   \delta\Bar{\Psi}(x)=\alpha\Bar{\Psi}(x)\gamma_5,
\label{eq:(3.8)}
\end{equation}
where $\alpha$ is an infinitesimal parameter. The second integral, which has
the form of a Jacobian under the change of integration variables
\begin{equation}
   \delta\psi(x)=\alpha\gamma_5\Psi(x),\qquad
   \delta\bar{\psi}(x)=\alpha\bar{\Psi}(x)\gamma_5,
\label{eq:(3.9)}
\end{equation}
is equal to the ``anomalous term'' in~Eq.~\eqref{eq:(2.11)}:
\begin{equation}
   -\int d^Dx\,
   \tr\left[
   \gamma_5\Psi(x')\frac{\overleftarrow{\delta}}{\delta\psi(x)}
   +\frac{\overrightarrow{\delta}}{\delta\Bar{\psi}(x)}\Bar{\Psi}(x')\gamma_5
   \right]
   =\int d^Dx\,
   \tr(-2i)\gamma_5\frac{\overrightarrow{\delta}}{\delta\Bar{\psi}(x')}
   S\frac{\overleftarrow{\delta}}{\delta\psi(x)}.
\label{eq:(3.10)}
\end{equation}

As we explained in~Sect.~\ref{sec:2.2}, the vanishing of~$Q_5$ is the condition
for global chiral invariance in the present formulation. If the anomalous
term~\eqref{eq:(3.10)} vanishes, this becomes the invariance of~$\mit\Gamma$
under the \emph{conventional} global chiral transformation:
\begin{equation}
   Q_5=\int d^Dx\,
   \left[
   {\mit\Gamma}\frac{\overleftarrow{\delta}}{\delta\Psi(x)}
   \gamma_5\Psi(x')
   +\Bar{\Psi}(x')\gamma_5
   \frac{\overrightarrow{\delta}}{\delta\Bar{\Psi}(x)}{\mit\Gamma}
   \right]=0.
\label{eq:(3.11)}
 \end{equation}
In~$D=4$ the axial vector current suffers from the axial anomaly, but if we
restrict ourselves only to the field configurations satisfying
\begin{equation}
   \int d^4x\,
   \epsilon_{\alpha\beta\gamma\delta}
   \mathcal{F}_{\alpha\beta}(x)\mathcal{F}_{\gamma\delta}(x)=0,
\label{eq:(3.12)}
\end{equation}
where
$\mathcal{F}_{\mu\nu}\equiv%
\partial_\mu\mathcal{A}_\nu-\partial_\nu\mathcal{A}_\mu$, the anomalous
term~\eqref{eq:(3.10)} still vanishes, and we can impose the global axial
invariance by~Eq.~\eqref{eq:(3.11)}.

Please note that if the Wilson action is bilinear in fermion fields
\begin{equation}
   S=\int d^Dx\,\int d^Dy\,\Bar{\psi}(x)iD(x,y)\psi(y),
\label{eq:(3.13)}
\end{equation}
the anomalous term becomes
\begin{equation}
   \int d^Dx\,
   \tr\left[
   -2i\gamma_5\frac{\overrightarrow{\delta}}{\delta\Bar{\psi}(x')}
   S\frac{\overleftarrow{\delta}}{\delta\psi(x)}\right]
   =2\int d^Dx\,\tr\gamma_5D(x,x).
\label{eq:(3.14)}
\end{equation}
This is an expression of the topological charge well known in lattice gauge
theory~\cite{Neuberger:1997fp,Hasenfratz:1997ft,Neuberger:1998wv,%
Hasenfratz:1998ri,Luscher:1998pqa,Hasenfratz:1998jp}. For computation of the
chiral anomaly in GFERG, see~Refs.~\cite{Miyakawa:2021hcx,Miyakawa:2022qbz}.
Our discussion of chiral invariance given in this subsection should be
considered preliminary; we would like to come back to the subject in a later
publication with a more detailed analysis.

\section{GFERG equation}
\label{sec:4}
In this section, we derive the GFERG flow equation for the 1PI
action~$\mit\Gamma$. For this, we first note that the 1PI action~$\mit\Gamma$
at each~$\tau$ is given by the Legendre transformation~\eqref{eq:(3.1)} of the
Wilson action~$S$ at~$\tau$. Since the Legendre transformation does not have
any explicit $\tau$-dependence, we have
\begin{equation}
   \partial_\tau{\mit\Gamma}=\partial_\tau S,
\label{eq:(4.1)}
\end{equation}
where the right-hand side is given by~Eq.~\eqref{eq:(2.1)}.

Next, we introduce 
\begin{subequations}\label{eq:(4.2)}
\begin{align}
   \left\llbracket\mathcal{A}_\mu(x')\Psi(x)\right\rrbracket
   &\equiv e^{-S}\Hat{s}A_\mu(x')\psi(x)\Hat{s}^{-1}e^S,
\\
   \left\llbracket\mathcal{A}_\mu(x')\mathcal{A}_\mu(x'')\Psi(x)\right\rrbracket
   &\equiv e^{-S}\Hat{s}A_\mu(x')A_\mu(x'')\psi(x)\Hat{s}^{-1}e^S,
\\
   \left\llbracket\mathcal{A}_\mu(x')\Bar{\Psi}(x)\right\rrbracket
   &\equiv e^{-S}\Hat{s}A_\mu(x')\Bar{\psi}(x)\Hat{s}^{-1}e^S,
\\
   \left\llbracket\mathcal{A}_\mu(x')\mathcal{A}_\mu(x'')\Bar{\Psi}(x)
   \right\rrbracket
   &\equiv e^{-S}\Hat{s}A_\mu(x')A_\mu(x'')\Bar{\psi}(x)\Hat{s}^{-1}e^S.
\end{align}
\end{subequations}
These define composite operators corresponding to products of fields. Note that
these appear in the GFERG equation~\eqref{eq:(2.1)}. With this in mind, we
introduce the following composite operators by combining scaling and diffusion:
\begin{subequations}\label{eq:(4.3)}
\begin{align}
   &\mathcal{O}_\mu(x)
   \equiv
   \left(\frac{D-2}{2}+\gamma+x\cdot\partial+2\partial^2\right)
   \mathcal{A}_\mu(x),
\\
   &\mathcal{O}_F(x)
\notag\\
   &\equiv
   \left(\frac{D-1}{2}+\gamma_F+x\cdot\partial+2\partial^2\right)\Psi(x)
   -4ie\left\llbracket\mathcal{A}_\mu(x')\partial_\mu\Psi(x)\right\rrbracket
   -2e^2\left\llbracket\mathcal{A}_\mu(x')\mathcal{A}_\mu(x'')\Psi(x)
   \right\rrbracket,
\\
   &\Bar{\mathcal{O}}_F(x)
\notag\\
   &\equiv
   \left(\frac{D-1}{2}+\gamma_F+x\cdot\partial+2\partial^2\right)\Bar{\Psi}(x)
   +4ie\left\llbracket\mathcal{A}_\mu(x')\partial_\mu\Bar{\Psi}(x)
   \right\rrbracket
   -2e^2\left\llbracket\mathcal{A}_\mu(x')\mathcal{A}_\mu(x'')\Bar{\Psi}(x)
   \right\rrbracket.
\end{align}
\end{subequations}

We then introduce the equation-of-motion composite operators by
\begin{subequations}\label{eq:(4.4)}
\begin{align}
   \mathcal{E}_A(x)
   &\equiv-e^{-S}\frac{\delta}{\delta A_\mu(x)}\left[e^S\mathcal{O}_\mu(x')
   \right],
\\
   \mathcal{E}_F(x)
   &\equiv\tr\left[e^S\mathcal{O}_F(x')\right]
   \frac{\overleftarrow{\delta}}{\delta\psi(x)}e^{-S},
\\
   \Bar{\mathcal{E}}_F(x)
   &\equiv e^{-S}\tr
   \frac{\overrightarrow{\delta}}{\delta\Bar{\psi}(x)}
   \left[e^S\Bar{\mathcal{O}}_F(x')\right],
\end{align}
\end{subequations}
so that the GFERG equation~\eqref{eq:(2.1)} can be expressed concisely as
\begin{equation}
   \partial_\tau{\mit\Gamma}
   =\int d^Dx\,
   \left[
   \mathcal{E}_A(x)+\mathcal{E}_F(x)+\Bar{\mathcal{E}}_F(x)
   \right].
\label{eq:(4.5)}
\end{equation}

Since $\mit\Gamma$ is a functional of $\mathcal{A}_\mu$, $\Psi$,
and~$\Bar{\Psi}$, we wish to express the right-hand side above in terms of
those field variables. We do this in steps.

First, it follows from definition~\eqref{eq:(4.4)} that
\begin{subequations}\label{eq:(4.6)}
\begin{align}
   \mathcal{E}_A(x)
   &=-\frac{\delta{\mit\Gamma}}{\delta\mathcal{A}_\mu(x)}\mathcal{O}_\mu(x')
   -\frac{\delta}{\delta A_\mu(x)}\mathcal{O}_\mu(x'),
\\
   \mathcal{E}_F(x)
   &=\tr\left[
   \mathcal{O}_F(x')\cdot
   {\mit\Gamma}\frac{\overleftarrow{\delta}}{\delta\Psi(x)}
   +\mathcal{O}_F(x')\frac{\overleftarrow{\delta}}{\delta\psi(x)}
   \right],
\\
   \Bar{\mathcal{E}}_F(x)
   &=\tr\left[
   \frac{\overrightarrow{\delta}}{\delta\Bar{\Psi}(x)}{\mit\Gamma}\cdot
   \Bar{\mathcal{O}}_F(x')
   +\frac{\overrightarrow{\delta}}{\delta\Bar{\psi}(x)}
   \Bar{\mathcal{O}}_F(x')
   \right],
\end{align}
\end{subequations}
where we have noted~Eq.~\eqref{eq:(3.3)}. $\mathcal{O}_\mu(x)$,
$\mathcal{O}_F(x)$, and~$\Bar{\mathcal{O}}_F(x')$ are given
by~Eq.~\eqref{eq:(4.3)} and they contain composite operators
in~Eq.~\eqref{eq:(4.2)}. The products are explicitly given by,
from~Eq.~\eqref{eq:(2.12)},
\begin{subequations}\label{eq:(4.7)}
\begin{align}
   \left\llbracket\mathcal{A}_\mu(x')\Psi(x)\right\rrbracket
   &=\mathcal{A}_\mu(x')\Psi(x)+\frac{\delta}{\delta A_\mu(x')}\Psi(x)
\notag\\
   &=\mathcal{A}_\mu(x')\Psi(x)
   +i\frac{\overrightarrow{\delta}}{\delta\Bar{\psi}(x)}\mathcal{A}_\mu(x'),
\\
   \left\llbracket\mathcal{A}_\mu(x')\Bar{\Psi}(x)\right\rrbracket
   &=\mathcal{A}_\mu(x')\Bar{\Psi}(x)
   +\frac{\delta}{\delta A_\mu(x')}\Bar{\Psi}(x)
\notag\\
   &=\mathcal{A}_\mu(x')\Bar{\Psi}(x)
   +\mathcal{A}_\mu(x')i\frac{\overleftarrow{\delta}}{\delta\psi(x)}.
\end{align}
\end{subequations}
The equality of the two expressions on the right-hand sides follows from the
definition~\eqref{eq:(4.2)}, because $A_\mu(x')\psi(x)=\psi(x)A_\mu(x')$ etc.
Similarly, we have
\begin{subequations}\label{eq:(4.8)}
\begin{align}
   \left\llbracket\mathcal{A}_\mu(x')\mathcal{A}_\mu(x'')\Psi(x)
   \right\rrbracket
   &=\left[
   \mathcal{A}_\mu(x')+\frac{\delta}{\delta A_\mu(x')}
   \right]
   \left\llbracket\mathcal{A}_\mu(x'')\Psi(x)\right\rrbracket
\notag\\
   &=\left[
   \mathcal{A}_\mu(x'')+\frac{\delta}{\delta A_\mu(x'')}
   \right]
   \left\llbracket\mathcal{A}_\mu(x')\Psi(x)\right\rrbracket
\notag\\
   &=\left[
   \Psi(x)+i\frac{\overrightarrow{\delta}}{\delta\Bar{\psi}(x)}
   \right]
   \left\llbracket\mathcal{A}_\mu(x')\mathcal{A}_\mu(x'')\right\rrbracket,
\\
   \left\llbracket\mathcal{A}_\mu(x')\mathcal{A}_\mu(x'')\Bar{\Psi}(x)
   \right\rrbracket
   &=\left[
   \mathcal{A}_\mu(x')+\frac{\delta}{\delta A_\mu(x')}
   \right]
   \left\llbracket\mathcal{A}_\mu(x'')\Bar{\Psi}(x)\right\rrbracket
\notag\\
   &=\left[
   \mathcal{A}_\mu(x'')+\frac{\delta}{\delta A_\mu(x'')}
   \right]
   \left\llbracket\mathcal{A}_\mu(x')\Bar{\Psi}(x)\right\rrbracket
\notag\\
   &=\left\llbracket\mathcal{A}_\mu(x')\mathcal{A}_\mu(x'')\right\rrbracket
   \left[
   \Bar{\Psi}(x)+i\frac{\overleftarrow{\delta}}{\delta\psi(x)}
   \right].
\end{align}
\end{subequations}

From these expressions, as the explicit forms of the equation-of-motion
composite operators in~Eq.~\eqref{eq:(4.6)}, we have
\begin{align}
   \mathcal{E}_A(x)
   &=-\frac{\delta{\mit\Gamma}}{\delta\mathcal{A}_\mu(x)}
   \left(2\partial^2+\frac{D-2}{2}+\gamma+x\cdot\partial\right)
   \mathcal{A}_\mu(x)
\notag\\
   &\qquad{}
   -\left(2\partial_{x'}^2+\frac{D-2}{2}+\gamma+x'\cdot\partial_{x'}\right)
   \frac{\delta\mathcal{A}_\mu(x')}{\delta A_\mu(x)},
\label{eq:(4.9)}
\end{align}
\begin{align}
   \mathcal{E}_F(x)
   &=-{\mit\Gamma}\frac{\overleftarrow{\delta}}{\delta\Psi(x)}
   \left(2\partial^2+\frac{D-1}{2}+\gamma_F+x\cdot\partial\right)
   \Psi(x)
\notag\\
   &\qquad{}
   +\tr\left(2\partial_{x'}^2+\frac{D-1}{2}+\gamma_F+x'\cdot\partial_{x'}\right)
   \Psi(x')\frac{\overleftarrow{\delta}}{\delta\psi(x)}
\notag\\
   &\qquad{}
   +4ie{\mit\Gamma}\frac{\overleftarrow{\delta}}{\delta\Psi(x)}
   \left[
   \mathcal{A}_\mu(x)\partial_\mu\Psi(x)
   +\partial_\mu\frac{\delta}{\delta A_\mu(x')}\Psi(x)
   \right]
\notag\\
   &\qquad{}
   -4ie\tr
   \left[
   \mathcal{A}_\mu(x')\partial_\mu\Psi(x')
   +\partial_\mu\frac{\delta}{\delta A_\mu(x'')}\Psi(x')
   \right]
   \frac{\overleftarrow{\delta}}{\delta\psi(x)}
\notag\\
   &\qquad{}
   +2e^2{\mit\Gamma}\frac{\overleftarrow{\delta}}{\delta\Psi(x)}
   \biggl\{
   \mathcal{A}_\mu(x)\mathcal{A}_\mu(x)\Psi(x)
   +\mathcal{A}_\mu(x)\frac{\delta}{\delta A_\mu(x')}\Psi(x)
\notag\\
   &\qquad\qquad\qquad\qquad\qquad{}
   +\frac{\delta}{\delta A_\mu(x')}
   \left[\mathcal{A}_\mu(x'')\Psi(x)\right]
   +\frac{\delta^2}{\delta A_\mu(x')\delta A_\mu(x'')}\Psi(x)
   \biggr\}
\notag\\
   &\qquad{}
   -2e^2\tr
   \biggl\{
   \mathcal{A}_\mu(x'')\mathcal{A}_\mu(x'')\Psi(x')
   +\mathcal{A}_\mu(x'')\frac{\delta}{\delta A_\mu(x'')}\Psi(x')
\notag\\
   &\qquad\qquad\qquad\qquad{}
   +\frac{\delta}{\delta A_\mu(x'')}
   \left[\mathcal{A}_\mu(x''')\Psi(x')\right]
   +\frac{\delta^2}{\delta A_\mu(x'')\delta A_\mu(x''')}\Psi(x')
   \biggr\}
   \frac{\overleftarrow{\delta}}{\delta\psi(x)},
\label{eq:(4.10)}
\end{align}
and
\begin{align}
   \Bar{\mathcal{E}}_F(x)
   &=-\left(2\partial^2+\frac{D-1}{2}+\gamma_F+x\cdot\partial\right)
   \Bar{\Psi}(x)\cdot
   \frac{\overrightarrow{\delta}}{\delta\Bar{\Psi}(x)}{\mit\Gamma}
\notag\\
   &\qquad{}
   +\tr\left(2\partial_{x'}^2+\frac{D-1}{2}+\gamma_F+x'\cdot\partial_{x'}\right)
   \frac{\overrightarrow{\delta}}{\delta\Bar{\psi}(x)}\Bar{\Psi}(x')
\notag\\
   &\qquad{}
   -4ie
   \left[
   \mathcal{A}_\mu(x)\partial_\mu\Bar{\Psi}(x)
   +\partial_\mu\frac{\delta}{\delta A_\mu(x')}\Bar{\Psi}(x)
   \right]
   \frac{\overrightarrow{\delta}}{\delta\Bar{\Psi}(x)}{\mit\Gamma}
\notag\\
   &\qquad{}
   +4ie\tr
   \frac{\overrightarrow{\delta}}{\delta\Bar{\psi}(x)}
   \left[
   \mathcal{A}_\mu(x')\partial_\mu\Bar{\Psi}(x')
   +\partial_\mu\frac{\delta}{\delta A_\mu(x'')}\Bar{\Psi}(x')
   \right]
\notag\\
   &\qquad{}
   +2e^2
   \biggl\{
   \mathcal{A}_\mu(x)\mathcal{A}_\mu(x)\Bar{\Psi}(x)
   +\mathcal{A}_\mu(x)\frac{\delta}{\delta A_\mu(x')}\Bar{\Psi}(x)
\notag\\
   &\qquad\qquad\qquad{}
   +\frac{\delta}{\delta A_\mu(x')}
   \left[\mathcal{A}_\mu(x'')\Bar{\Psi}(x)\right]
   +\frac{\delta^2}{\delta A_\mu(x')\delta A_\mu(x'')}\Bar{\Psi}(x)
   \biggr\}
   \frac{\overrightarrow{\delta}}{\delta\Bar{\Psi}(x)}{\mit\Gamma}
\notag\\
   &\qquad{}
   -2e^2\tr
   \frac{\overrightarrow{\delta}}{\delta\Bar{\psi}(x)}
   \biggl\{
   \mathcal{A}_\mu(x'')\mathcal{A}_\mu(x'')\Bar{\Psi}(x')
   +\mathcal{A}_\mu(x'')\frac{\delta}{\delta A_\mu(x'')}\Bar{\Psi}(x')
\notag\\
   &\qquad\qquad\qquad\qquad\qquad{}
   +\frac{\delta}{\delta A_\mu(x'')}
   \left[\mathcal{A}_\mu(x''')\Bar{\Psi}(x')\right]
   +\frac{\delta^2}{\delta A_\mu(x'')\delta A_\mu(x''')}\Bar{\Psi}(x')
   \biggr\}.
\label{eq:(4.11)}
\end{align}

These expressions contain the functional derivatives of the Legendre
transformed variables $(\mathcal{A}_\mu,\Psi,\Bar{\Psi})$ with respect to the
original field variables $(A_\mu,\psi,\Bar{\psi})$. By the chain rule of
differentiation, the first derivatives satisfy
\begin{align}
   &\begin{pmatrix}
   \delta_{\mu\nu}&0&0\\
   0&1&0\\
   0&0&1\\
   \end{pmatrix}\delta(x-y)
   =\begin{pmatrix}
   \frac{\delta A_\nu(y)}{\delta A_\mu(x)}&
   \frac{\delta\Bar{\psi}(y)}{\delta A_\mu(x)}&
   \frac{\delta\psi(y)}{\delta A_\mu(x)}\\
   \frac{\overrightarrow{\delta}}{\delta\Bar{\psi}(x)}A_\nu(y)&
   \frac{\overrightarrow{\delta}}{\delta\Bar{\psi}(x)}\Bar{\psi}(y)&
   \frac{\overrightarrow{\delta}}{\delta\Bar{\psi}(x)}\psi(y)\\
   \frac{\overrightarrow{\delta}}{\delta\psi(x)}A_\nu(y)&
   \frac{\overrightarrow{\delta}}{\delta\psi(x)}\Bar{\psi}(y)&
   \frac{\overrightarrow{\delta}}{\delta\psi(x)}\psi(y)\\
   \end{pmatrix}
\notag\\
   &=\int d^Dz\,
   \begin{pmatrix}
   \frac{\delta\mathcal{A}_\rho(z)}{\delta A_\mu(x)}&
   \frac{\delta\Bar{\Psi}(z)}{\delta A_\mu(x)}&
   \frac{\delta\Psi(z)}{\delta A_\mu(x)}\\
   \frac{\overrightarrow{\delta}}{\delta\Bar{\psi}(x)}\mathcal{A}_\rho(z)&
   \frac{\overrightarrow{\delta}}{\delta\Bar{\psi}(x)}\Bar{\Psi}(z)&
   \frac{\overrightarrow{\delta}}{\delta\Bar{\psi}(x)}\Psi(z)\\
   \frac{\overrightarrow{\delta}}{\delta\psi(x)}\mathcal{A}_\rho(z)&
   \frac{\overrightarrow{\delta}}{\delta\psi(x)}\Bar{\Psi}(z)&
   \frac{\overrightarrow{\delta}}{\delta\psi(x)}\Psi(z)\\
   \end{pmatrix}
   \begin{pmatrix}
   \frac{\delta A_\nu(y)}{\delta\mathcal{A}_\rho(z)}&
   \frac{\delta\Bar{\psi}(y)}{\delta\mathcal{A}_\rho(z)}&
   \frac{\delta\psi(y)}{\delta\mathcal{A}_\rho(z)}\\
   \frac{\overrightarrow{\delta}}{\delta\Bar{\Psi}(z)}A_\nu(y)&
   \frac{\overrightarrow{\delta}}{\delta\Bar{\Psi}(z)}\Bar{\psi}(y)&
   \frac{\overrightarrow{\delta}}{\delta\Bar{\Psi}(z)}\psi(y)\\
   \frac{\overrightarrow{\delta}}{\delta\Psi(z)}A_\nu(y)&
   \frac{\overrightarrow{\delta}}{\delta\Psi(z)}\Bar{\psi}(y)&
   \frac{\overrightarrow{\delta}}{\delta\Psi(z)}\psi(y)\\
   \end{pmatrix}.
\label{eq:(4.12)}
\end{align}
Thus, the first derivatives are given by the inverse of
\begin{align}
   &\begin{pmatrix}
   \frac{\delta A_\nu(y)}{\delta\mathcal{A}_\mu(x)}&
   \frac{\delta\Bar{\psi}(y)}{\delta\mathcal{A}_\mu(x)}&
   \frac{\delta\psi(y)}{\delta\mathcal{A}_\mu(x)}\\
   \frac{\overrightarrow{\delta}}{\delta\Bar{\Psi}(x)}A_\nu(y)&
   \frac{\overrightarrow{\delta}}{\delta\Bar{\Psi}(x)}\Bar{\psi}(y)&
   \frac{\overrightarrow{\delta}}{\delta\Bar{\Psi}(x)}\psi(y)\\
   \frac{\overrightarrow{\delta}}{\delta\Psi(x)}A_\nu(y)&
   \frac{\overrightarrow{\delta}}{\delta\Psi(x)}\Bar{\psi}(y)&
   \frac{\overrightarrow{\delta}}{\delta\Psi(x)}\psi(y)\\
   \end{pmatrix}
\notag\\
   &=\begin{pmatrix}
   \delta_{\mu\nu}\delta(x-y)
   -\dfrac{\delta^2{\mit\Gamma}}
   {\delta\mathcal{A}_\mu(x)\delta\mathcal{A}_\nu(y)}&
   -i\dfrac{\delta}{\delta\mathcal{A}_\mu(x)}{\mit\Gamma}
   \dfrac{\overleftarrow{\delta}}{\delta\Psi(y)}&
   -i\dfrac{\delta}{\delta\mathcal{A}_\mu(x)}
   \dfrac{\overrightarrow{\delta}}{\delta\Bar{\Psi}(y)}{\mit\Gamma}\\
   -\dfrac{\overrightarrow{\delta}}{\delta\Bar{\Psi}(x)}
   \dfrac{\delta{\mit\Gamma}}{\delta\mathcal{A}_\nu(y)}&
   \delta(x-y)-i\dfrac{\overrightarrow{\delta}}{\delta\Bar{\Psi}(x)}
   {\mit\Gamma}\dfrac{\overleftarrow{\delta}}{\delta\Psi(y)}&
   -i\dfrac{\overrightarrow{\delta}}{\delta\Bar{\Psi}(x)}
   \dfrac{\overrightarrow{\delta}}{\delta\Bar{\Psi}(y)}{\mit\Gamma}\\
   \dfrac{\delta{\mit\Gamma}}{\delta\mathcal{A}_\nu(y)}
   \dfrac{\overleftarrow{\delta}}{\delta\Psi(x)}&
   i{\mit\Gamma}\dfrac{\overleftarrow{\delta}}{\delta\Psi(x)}
   \dfrac{\overleftarrow{\delta}}{\delta\Psi(y)}&
   \delta(x-y)-i
   \dfrac{\overrightarrow{\delta}}{\delta\Bar{\Psi}(y)}{\mit\Gamma}
   \dfrac{\overleftarrow{\delta}}{\delta\Psi(x)}\\
   \end{pmatrix}.
\label{eq:(4.13)}
\end{align}
where we have used~Eq.~\eqref{eq:(3.2)}. These first-order functional
derivatives are common in the ordinary ERG formulation for the 1PI action. What
is peculiar to the GFERG formulation is the presence of the second- and
third-order functional derivatives appearing
in~$\mathcal{E}_F(x)$~\eqref{eq:(4.10)}
and~$\Bar{\mathcal{E}}_F(x)$~\eqref{eq:(4.11)}. These higher-order derivatives
are necessary for the manifest gauge invariance of~$\mit\Gamma$, and they can
be obtained by differentiating further the elements of the above inverse
matrix.

This extra labor is required for the sake of manifest gauge invariance for the
1PI Wilson action.

\section{Perturbative solution}
\label{sec:5}
If we restrict ourselves only to the Wilson action parametrized by the gauge
coupling~$e$, the electron mass~$m$, and the gauge-fixing parameter~$\xi$, we
can solve the GFERG equation~\eqref{eq:(4.5)} as a power series in~$e$, where
the lowest-order term is the Gaussian fixed point. We may directly solve
Eq.~\eqref{eq:(4.5)} or apply the Legendre transformation~\eqref{eq:(3.1)} to
the Wilson action~$S$ that has been obtained perturbatively to order~$e^2$
in~Ref.~\cite{Miyakawa:2021wus}. Following the latter approach, a
straightforward calculation gives
\begin{align}
   {\mit\Gamma}
   &=-\frac{1}{2}\int_k\,
   e^{k^2}\mathcal{A}_\mu(k)e^{k^2}\mathcal{A}_\nu(-k)
   \left\{
   (\delta_{\mu\nu}k^2-k_\mu k_\nu)
   \left[1-e^2\frac{\widetilde{V}_T(k)}{k^2}\right]
   +\frac{1}{\xi}k_\mu k_\nu\right\}
\notag\\
   &\qquad{}
   -\int_p\,\Bar{\Psi}(-p)e^{p^2}
   \left[(\Slash{p}+im)-e^2\widetilde{V}_F(p)\right]
   e^{p^2}\Psi(p)
\notag\\
   &\qquad{}
   +e\int_{p,k}\,
   \Bar{\Psi}(-p-k)e^{(p+k)^2}
   \widetilde{V}_\mu(p,k)
   e^{p^2}\Psi(p)e^{k^2}\mathcal{A}_\mu(k)
\notag\\
   &\qquad{}
   +e^2\int_{p,k}\,
   \Bar{\Psi}(-p-k-l)e^{(p+k+l)^2}\overline{V}_{\mu\nu}(p,k,l)
   e^{p^2}\Psi(p)e^{k^2}\mathcal{A}_\mu(k)e^{l^2}\mathcal{A}_\nu(l)
\notag\\
   &\qquad{}+O(e^3),
\label{eq:(5.1)}
\end{align}
where $\overline{V}_{\mu\nu}(p,k,l)=\overline{V}_{\nu\mu}(p,l,k)$ is symmetric.
The functions $\widetilde{V}_\mu(p,k)$ and~$\overline{V}_{\mu\nu}(p,k,l)$ are
given in~Ref.~\cite{Miyakawa:2021wus} as follows:
\begin{align}
   \widetilde{V}_\mu(p,k) 
   &=\gamma_\mu+2(\Slash{p}+\Slash{k}+im)p_\mu F((p+k)^2-p^2-k^2)
\notag\\
   &\qquad{}
   +2(\Slash{p}+im)(p+k)_\mu F(p^2-(p+k)^2-k^2),
\label{eq:(5.2)}
\end{align}
where
\begin{equation}
   F(x)\equiv\frac{e^x-1}{x}
\label{eq:(5.3)}
\end{equation}
and
\begin{align}
   \overline{V}_{\mu\nu}(p,k,l)
   &=-\delta_{\mu\nu}
   \bigl[
   (\Slash{p}+\Slash{k}+\Slash{l}+im)F((p+k+l)^2-p^2-k^2-l^2)
\notag\\
   &\qquad\qquad{}
   +(\Slash{p}+im)F(p^2-(p+k+l)^2-k^2-l^2)
   \bigr]
\notag\\
   &\qquad{}
   -4X_{\mu\nu}(p,k,l),
\label{eq:(5.4)}
\end{align}
where
\begin{align}
   &X_{\mu\nu}(p,k,l)=X_{\nu\mu}(p,l,k)
\notag\\
   &=\frac{1}{4}\gamma_\mu p_\nu F((p+l)^2-p^2-l^2)
   +\frac{1}{4}(p+k+l)_\mu\gamma_\nu F((p+l)^2-(p+k+l)^2-k^2)
\notag\\
   &\qquad{}
   +\frac{1}{2}(\Slash{p}+\Slash{l}+im)(p+k+l)_\mu p_\nu
   F((p+l)^2-(p+k+l)^2-k^2)F((p+l)^2-p^2-l^2)
\notag\\
   &\qquad{}
   +\frac{1}{2}\frac{(\Slash{p}+\Slash{k}+\Slash{l}+im)(p+l)_\mu p_\nu}
   {(p+k+l)^2-(p+l)^2-k^2}
\notag\\
   &\qquad\qquad{}
   \times
   \left[F((p+k+l)^2-p^2-k^2-l^2)-F((p+l)^2-p^2-l^2)\right]
\notag\\
   &\qquad{}
   +\frac{1}{2}\frac{(\Slash{p}+im)(p+k+l)_\mu(p+l)_\nu}{p^2-(p+l)^2-l^2}
\notag\\
   &\qquad\qquad{}
   \times
   \left[F(p^2-(p+k+l)^2-k^2-l^2)-F((p+l)^2-(p+k+l)^2-k^2)\right]
\notag\\
   &\quad{}
   +(\mu\leftrightarrow \nu,k\leftrightarrow l).
\label{eq:(5.5)}
\end{align}
As for $\widetilde{V}_T(k)$ and~$\widetilde{V}_F(p)$ in~Eq.~\eqref{eq:(5.1)}
(one-loop corrections to the kinetic terms), only the leading terms in the
momentum expansions have been explicitly computed
in~Ref.~\cite{Miyakawa:2021wus}.

The above results, especially the explicit form of the
function~$X_{\mu\nu}(p,k,l)$~\eqref{eq:(5.5)}, might appear complicated, but the
basic structure of the 1PI action~\eqref{eq:(5.1)} is much simpler than that of
the corresponding Wilson action~$S$. For example, the Wilson action~$S$
contains a term such as~\cite{Miyakawa:2021wus}
\begin{equation}
   \frac{1}{2}e^2\int_{p,q,k}
   \Bar{\Psi}(-p-k)e^{(p+k)^2}\widetilde{V}_\mu(p,k)e^{p^2}\Psi(p)
   \Bar{\Psi}(-q)e^{q^2}\widetilde{V}_\nu(q+k,-k)e^{(q+k)^2}\Psi(q+k)
   h_{\mu\nu}(k),
\label{eq:(5.6)}
\end{equation}
where
\begin{equation}
   h_{\mu\nu}(k)
   =\left(\delta_{\mu\nu}-\frac{k_\mu k_\nu}{k^2}\right)\frac{1}{e^{-2k^2}+k^2}
   +\frac{k_\mu k_\nu}{k^2}\frac{\xi}{\xi e^{-2k^2}+k^2}
\label{eq:(5.7)}
\end{equation}
is the high-momentum or short-distance propagator of the photon. This
four-Fermi term, which may be regarded as a one-particle \emph{reducible\/}
(1PR) part, is removed from the 1PI action~$\mit\Gamma$~\eqref{eq:(5.1)} after
the Legendre transformation. Similarly, $\overline{V}_{\mu\nu}(p,k,l)$ given
by~Eq.\eqref{eq:(5.4)} is obtained from~$\widetilde{V}_{\mu\nu}(p,k,l)$ given
in~Ref.~\cite{Miyakawa:2021wus} by the removal of 1PR terms.

Excluding the gauge-fixing term, $\mit\Gamma$ given by~Eq.~\eqref{eq:(5.1)}
should be invariant under the gauge transformation~\eqref{eq:(3.4)}. This
requires
\begin{equation}
   k_\mu\widetilde{V}_\mu(p,k)
   =e^{(p+k)^2-p^2-k^2}(\Slash{p}+\Slash{k}+im)
   -e^{p^2-(p+k)^2-k^2}(\Slash{p}+im),
\label{eq:(5.8)}
\end{equation}
and
\begin{equation}
   2k_\mu
   \overline{V}_{\mu\nu}(p,k,l)
   =e^{(p+k)^2-(p+k+l)^2-k^2}\widetilde{V}_\nu(p,l)
   -e^{(p+k)^2-p^2-k^2}\widetilde{V}_\nu(p+k,l).
\label{eq:(5.9)}
\end{equation}
It is easy to check that $\widetilde{V}_\mu$ given by~Eq.~\eqref{eq:(5.2)}
satisfies Eq.~\eqref{eq:(5.8)}. It is tedious but also possible to
verify~Eq.~\eqref{eq:(5.9)}. Thus, the 1PI Wilson action is gauge invariant.
Actually, with the exception of the gauge-fixing term, the right-hand side
of~Eq.~\eqref{eq:(5.1)} can be obtained by expanding the manifestly gauge
invariant expression\footnote{Here, we omit terms containing
$\widetilde{V}_T(k)$ and~$\widetilde{V}_F(p)$ for simplicity.}
\begin{equation}
   -\frac{1}{4}\int d^Dx\,
   \left[
   \partial_\mu\mathcal{A}_{-1\nu}(x)-\partial_\nu\mathcal{A}_{-1\mu}(x)
   \right]^2
   +i\int d^Dx\,
   \Bar{\Psi}_{-1}(x)
   \left[\Slash{\partial}-ie\Slash{\mathcal{A}}_{-1}(x)-m\right]
   \Psi_{-1}(x),
\label{eq:(5.10)}
\end{equation}
up to order~$e^2$. Here, the fields with subscript~$t=-1$ are the solutions of
the diffusion equations
\begin{subequations}\label{eq:(5.11)}
\begin{align}
   \partial_t\mathcal{A}_{t\mu}(x)
   &=\partial_\nu
   \left[
   \partial_\nu\mathcal{A}_{t\mu}(x)-\partial_\mu\mathcal{A}_{t\nu}(x)
   \right]
   +\alpha_0\partial_\mu\partial_\nu\mathcal{A}_{t\nu}(x),&
   \mathcal{A}_{0\mu}(x)&\equiv\mathcal{A}_\mu(x)
\\
   \partial_t\Psi_t(x)
   &=\left\{
   \left[\partial_\mu-ie\mathcal{A}_{t\mu}(x)\right]^2
   +ie\alpha_0\left[\partial_\mu\mathcal{A}_{t\mu}(x)\right]
   \right\}\Psi_t(x),&
   \Psi_0(x)&\equiv\Psi(x)
\\
   \partial_t\Bar{\Psi}_t(x)
   &=\left\{
   \left[\partial_\mu+ie\mathcal{A}_{t\mu}(x)\right]^2
   -ie\alpha_0\left[\partial_\mu\mathcal{A}_{t\mu}(x)\right]
   \right\}\Bar{\Psi}_t(x),&
   \Bar{\Psi}_0(x)&\equiv\Bar{\Psi}(x)
\end{align}
\end{subequations}
solved \emph{backward\/} from~$t=0$ to~$t=-1$. Our GFERG
equation~\eqref{eq:(2.1)} is based on the above diffusion equations
with~$\alpha_0=1$ (see, for instance, Eq.~(2.16)
of~Ref.~\cite{Miyakawa:2021wus}). These are the flow equations introduced
in~Refs.~\cite{Luscher:2010iy,Luscher:2013cpa}; although the solution
$(\mathcal{A}_{t\mu}(x),\Psi_t(x),\Bar{\Psi}_t(x))$ does not transform gauge
covariantly under~Eq.~\eqref{eq:(3.4)} for~$\alpha_0\neq0$, any gauge invariant
combination such as~Eq.~\eqref{eq:(5.10)} has been shown independent
of~$\alpha_0$~\cite{Luscher:2010iy}, and thus is invariant
under~Eq.~\eqref{eq:(3.4)}.\footnote{Under the gauge
transformation~\eqref{eq:(3.4)},
$(\mathcal{A}_{t\mu}(x),\Psi_t(x),\Bar{\Psi}_t(x))$ transforms
as~Eq.~\eqref{eq:(3.4)}, where $\chi(x)$ is replaced
by~$\chi_t(x)=e^{t\alpha_0\partial^2}\chi(x)$. This also explains the gauge
invariance of~Eq.~\eqref{eq:(5.10)}.}

\section{Conclusion}
\label{sec:6}
In this paper we have introduced a one-particle irreducible Wilson
action~$\mit\Gamma$ for QED using the GFERG (gradient flow exact
renormalization group) formalism. This is a straightforward extension of the
previous work~\cite{Miyakawa:2021wus} where the GFERG formalism was applied to
the construction of a Wilson action for QED. Realization of gauge invariance
within the ERG formalism has been of interest for its potential applications to
non-perturbative aspects of gauge theory, such as the search for non-trivial
fixed points of the renormalization group flow. But the gauge invariance is
modified within the ERG formalism in a non-trivial way, and it has been
difficult to ensure gauge invariance in practical calculations. We have
succeeded in constructing $\mit\Gamma$ with manifest invariance under the
\emph{conventional} gauge transformation, even though the GFERG flow equation
that preserves the gauge invariance is admittedly somewhat complicated.

There are three lines of development that we can pursue after this work. The
first is to find the relation, perhaps equivalence, of the GFERG formalism to
the ERG formalism. Do they share the same, i.e., physically equivalent,
fixed points? Do they give the same universality classes? We believe that the
two formalisms are physically equivalent, but the equivalence must be proven.

The second is more ambitious. We have referred to non-perturbative applications
a couple of times in this paper. There are already quite a few preceding works
that investigate non-perturbatively fixed points and critical exponents of
Abelian gauge theory~\cite{Aoki:1996fh,Gies:2004hy,Igarashi:2016gcf,%
Gies:2020xuh}. It should be quite interesting to reexamine the results using
our manifestly gauge invariant formalism.

Finally, we would like to generalize the present formulation to gauge-fixed
non-Abelian gauge theory. We suspect that it would be important to figure out
how to deal with Faddeev--Popov ghost fields and the Nakanishi--Lautrup
auxiliary field. It would be nice to have $\mit\Gamma$ manifestly invariant
under the \emph{conventional} gauge (or BRST) transformation.

\section*{Acknowledgements}
This work was partially supported by the Japan Society for the Promotion of
Science (JSPS) Grant-in-Aid for Scientific Research Grant Number JP20H01903.

\appendix

\section{BRST invariant 1PI action}
\label{sec:A}
We have shown that ${\mit\Gamma}[\mathcal{A}_\mu,\Psi,\Bar{\Psi}]$ defined
by~Eq.~\eqref{eq:(3.1)} transforms as~Eq.~\eqref{eq:(3.5)} under the gauge
transformation~\eqref{eq:(3.4)}. Now, the ghost action is given
by~Eq.~\eqref{eq:(2.5)}. With
\begin{align}
   C(x)
   &\equiv c(x)
   +\frac{\overrightarrow{\delta}}{\delta\Bar{c}(x)}S_{\text{ghost}}
   =\frac{E(-e^{-2\tau}\partial^2)e^{2\partial^2}}
   {E(-e^{-2\tau}\partial^2)e^{2\partial^2}-\partial^2}c(x),
\label{eq:(A1)}\\
   \Bar{C}(x)
   &\equiv\Bar{c}(x)
   +S_{\text{ghost}}\frac{\overleftarrow{\delta}}{\delta c(x)}
   =\frac{E(-e^{-2\tau}\partial^2)e^{2\partial^2}}
   {E(-e^{-2\tau}\partial^2)e^{2\partial^2}-\partial^2}\Bar{c}(x),
\label{eq:(A2)}
\end{align}
we define ${\mit\Gamma}_{\text{ghost}}[C,\Bar{C}]$ by
\begin{equation}
   {\mit\Gamma}_{\text{ghost}}
   -\int d^Dx\,\Bar{C}(x)C(x)
   =S_{\text{ghost}}
   +\int d^Dx\,\Bar{c}(x)c(x)
   -\int d^D x\,\left[\Bar{C}(x)c(x)+\Bar{c}(x)C(x)\right].
\label{eq:(A3)}
\end{equation}
This gives
\begin{equation}
   {\mit\Gamma}_{\text{ghost}}[C,\Bar{C}]
   =\int d^Dx\,\Bar{C}(x)
   \frac{\partial^2}{E(-e^{-2\tau}\partial^2)e^{2\partial^2}}C(x).
\label{eq:(A4)}
\end{equation}

To obtain a BRST transformation we choose
\begin{equation}
   \chi(x)=\eta C(x),
\label{eq:(A5)}
\end{equation}
where $\eta$ is an anticommuting constant, and transform the ghost fields by
\begin{equation}
   \delta C(x)=0,\qquad
   \delta\Bar{C}(x)=\frac{1}{\xi}\eta\partial\cdot\mathcal{A}(x).
\label{eq:(A6)}
\end{equation}
We then obtain
\begin{align}
   \delta{\mit\Gamma}
   &=-\int d^Dx\,\frac{1}{\xi}\partial^2\eta C(x)
   \frac{1}{E(-e^{-2\tau}e^{2\partial^2})e^{2\partial^2}}
   \partial\cdot\mathcal{A}(x),
\label{eq:(A7)}\\
   \delta{\mit\Gamma}_{\text{ghost}}
   &=\int d^Dx\,\frac{1}{\xi}\eta\partial\cdot\mathcal{A}(x)
   \frac{\partial^2}{E(-e^{-2\tau}\partial^2)e^{2\partial^2}}C(x).
\label{eq:(A8)}
\end{align}
Hence, the total 1PI Wilson action is BRST invariant:
\begin{equation}
   \delta({\mit\Gamma}+{\mit\Gamma}_{\text{ghost}})=0.
\label{eq:(A9)}
\end{equation}



%\bibliographystyle{ptephy}
%\bibliography{sample}
%
% once the .bbl file has been generated then place the text in your article.

%% \vspace{0.2cm}
%% \noindent
%% For references, note how to include DOI information from examples below. 

%This is added by T. Yoneya (editor-in-chief) on 2020/07/09.

\let\doi\relax

%without this code before the command "\begin{thebibliography}{}" , an error will be %flagged. When the bibliography is provided as separate .bib file, then this code %should be placed above the commands "\bibliographystyle{}" and "\bibliography{}" %inside the main TeX file. 

%% \begin{thebibliography}{9}

%% \bibitem{1}
%% J. P.~Blaizot, and E.~Iancu, Phys. Rep. {\bf 359}, 355 (2002).
%% \doi{https://doi.org/10.1016/S0370-1573(01)00061-8}

%% \bibitem{2}
%% M.~Gyulassy, and L.~McLerran, Nucl.\ Phys.\  A {\bf 750}, 30 (2005). \\ \doi{https://doi.org/10.1016/j.nuclphysa.2004.10.034}

%% \bibitem{3}
%% S.~Aoki et al. [JLQCD Collaboration], Phys. Rev. D 72, 054510 (2005). \\
%% \doi{https://doi.org/10.1103/PhysRevD.72.05451}

%% \bibitem{4}
%% S.~Alekhin, A.~Djouadi, and S.~Moch, Phys. Lett. B 716, 214 (2012) [arXiv:1207.0980 [hep-ph]]. \doi{https://doi.org/10.1016/j.physletb.2012.08.024}

%% \end{thebibliography}

\bibliographystyle{ptephy}
\bibliography{GFERG_1PI}

\end{document}